\documentclass[%
 reprint,
 amsmath,amssymb,
 aps,
]{revtex4-2}
\usepackage{multirow}
\usepackage{makecell}
\usepackage{graphicx}
\usepackage{dcolumn}
\usepackage{bm}

\begin{document}

\preprint{APS/123-QED}

\title{Study of two-electron one-photon transition produced in collision of $Ne^{6+}$ ions with Al target at low energies}

\author{Shashank Singh}
\affiliation{Department of Physics, Panjab University, Chandigarh-160014, India.}
\author{Mumtaz Oswal}%

\affiliation{%
 Department of Physics, DAV College, Sector 10, Chandigarh-160011, India.
}%

\author{K. P. Singh}
\affiliation{Department of Physics, Panjab University, Chandigarh-160014, India.
}%
\author{D. K. Swami}
\affiliation{%
Inter-University Accelerator Centre, Aruna Asaf Ali Marg, Near Vasant Kunj, New Delhi-110067, India.}%
\author{T. Nandi}
\email{nanditapan@gmail.com}
\affiliation{1003 Regal,
Mapsko Royal Ville, Sector-82, Gurgaon-122004, India.}
\altaffiliation{superannuated from
Inter-University Accelerator Centre, Aruna Asaf Ali Marg, Near Vasant Kunj, New Delhi-110067, India.}%


\date{\today}

\begin{abstract}
Two-electron one-photon transitions have been successfully observed for the Ne projectile and Al target at low energy regime. Experimental energy values of two-electron one-photon transitions are compared with previously reported theoretical and experimental values. Ionization cross-section of two-electron one-photon transition is reported.   
\end{abstract}

\maketitle


\section{Introduction}
\begin{figure}
\includegraphics[width=80mm,height=85mm,scale=01.0]{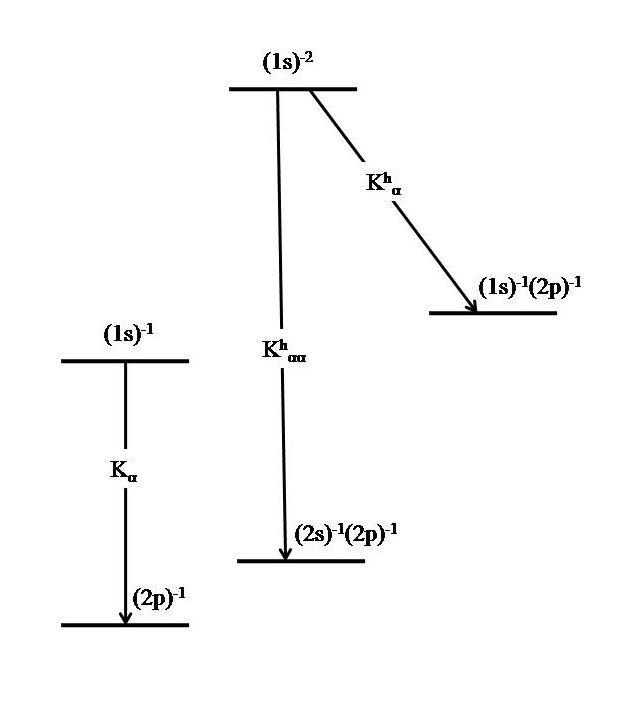}
\caption{\label{fig:1} Schematic diagram of the two electron one photon transition}
\end{figure}

In the heavy ion-atom collision processes, multiple ionization of different  shells created simultaneously is a common phenomenon. For a particular case, K shell of the target or projectile may be doubly ionized during ion-atom collision. Usually, these two vacancies of the K shell is filled by the subsequent electron of the higher shell and as a resultant, characteristics x-rays or Auger electron are emitted. Normally two vacancies are filled in two steps, firstly a hypersatellite line is emitted and secondly a satellite x-ray line as pictorially shown in Fig. \ref{fig:1}. Nevertheless, according to Heisenberg \cite{heisenberg1985quantentheorie}, Condon \cite{condon1930theory}, and Goudsmit and Gropper \cite{goudsmit1931many}, it is possible that both the vacancies of the K shell may be filled by a simultaneous transition of two electrons of upper shells. The simultaneous transition of two electrons to fill both K shell vacancies leads emission of one photon of energy of marginally higher than the double of the K shell transition energy. The energy of the two electron one photon (TEOP) transition (also called cooperative transition - $K^{-2}-L^{-2}$) can be estimated as the sum of the transition energies of hypersatellite ($K^{-2}-K^{-1}L^{-1}$) and satellite transition ($K^{-1}L^{-1}-L^{-2}$). In this estimation the vacancy in M or higher shells can be ignored as the their effect is very small. The schematic diagram of the TEOP transition is shown in the Fig. \ref{fig:1}.
\par
Firstly, \citet{wolfli1975two} reported two electron one photon transition for Ni-Ni and Ni-Fe systems at beam energies $\approx$ 0.8 MeV/amu. \citet{aaberg1976origin} showing that the two electron and one photon transition are due to the $1s^{-2}\rightarrow2s^{-1}2p^{-1}$ electric dipole transition by calculations of Hartee-Fock energy and intensity. Various authors \citet{wolfli1976calculation}, \citet{nagel1976two}, \cite{aaberg1976origin} and \citet{knudson1976energies}  reported the theoretical prediction of TEOP transition energies in different systems to compare observed energies. In which Hartee-Fock calculations have shown consistent result with the observed transition energies. \citet{stoller1977two} reported the TEOP transitions for Al-Al, O-Ca, Ca-Ca, Fe-Fe, Fe-Ni, Ni-Fe and Ni-Ni systems at the beam energies between 24 to 40 MeV. In this study, cross-sections of characteristic transitions and TEOP transitions were determined. Also, branching ratios between one and two electron transitions in doubly ionized K shell were determined to compare with various theoretical predictions and found in qualitative agreement. Besides th ion-atom collisions, \citet{auerhammer1988two} studied the TEOP transition for the Al target with electrons of energy 20 keV using high resolution crystal spectrometer. 
\par \citet{mukherjee1990branching} evaluated the branching ratio between one-electron one photon (OEOP) and TEOP transition in doubly ionized K shell using an equivalent two-particle model with wave functions including angular correlations and relaxation in the length and velocity gauges. \citet{mukherjee1997two} reported the theoretical estimation of excitation energies and transition probabilities for TEOP transition for the inner shell ionised atoms for Ne, Na, Mg, Al, Si, P, S, Cl and Ar. Also, the transition data presented for ${K^{S}_{\alpha \alpha}}$ transition ($2s^12p^1 \rightarrow 1s^2$) of the highly stripped He like ions $Ne^{8+}$, $Na^{9+}$, $Mg^{10+}$, $Al^{11+}$, $Si^{12+}$, $P^{13+}$, $S^{14+}$, $Cl^{15+}$ and $Ar^{16+}$. \citet{hoszowska2011first} observed first time TEOP transition in Mg, Al and Si by monochromatized synchrotron radiation  using a wavelength dispersive spectrometer. \citet{kasthurirangan2013observation} reported first time observation of the fluorescence-active doubly exited states in He-like Si, S and Cl ions. Both the transitions OEOP and TEOP are very sensitive to Breit interaction, quantum electrodynamics (QED), and
electron correlations \cite{diamant2000cu,briand1976two,kadrekar2010two}. 
\par 
The study of TEOP at low energy heavy ions is very few in literature and study of it is very important for highly stripped ions of He sequence \cite{andriamonje1991two,mukherjee1995interpretation} as these ions present in solar corona, flares and laboratory tokamak plasmas. So the less availability of literature at low energy heavy ions and importance of TEOP in various fields makes the TEOP study of high demand.    
 
\section{Experimental details}
\begin{figure}
\centering
\includegraphics[width=75mm,scale=01.0]{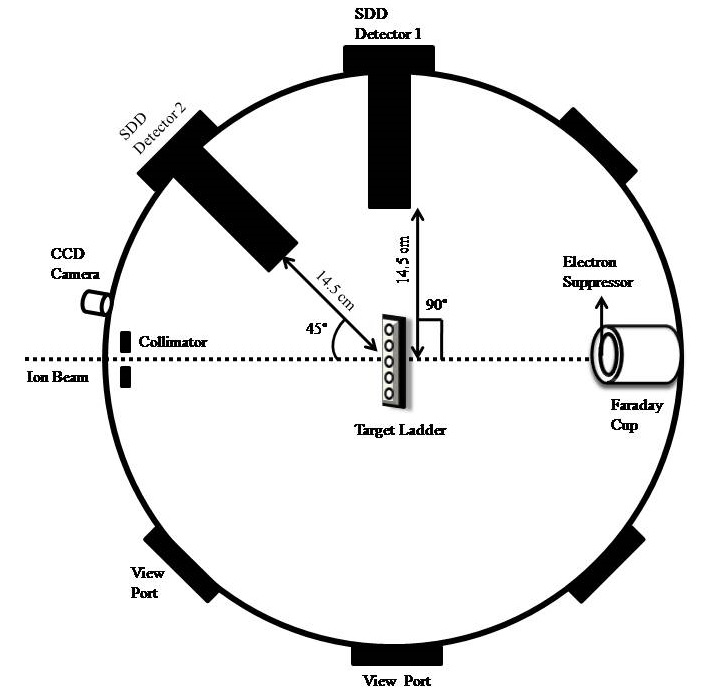}
\caption{\label{fig:2} Schematic diagram of the x-ray spectroscopy setup at LEIBF facility, IUAC, New Delhi}.
\end{figure}

This experiment was performed at LEIBF facility, IUAC, New Delhi. In  this experiment, we took Ne beam of 6+ charge state (1.8 and 2.1 MeV) from ECR ion source placed on a high voltage deck of 400 kV to bombard the Al target. The atomic physics chamber is situated at $75^\circ$ beam line of LEIBF facility. The schematic diagram of scattering chamber with experimental setup is shown in Fig. \ref{fig:2}. High vacuum was maintained inside the chamber using turbo-molecular pump. To measure the x-rays, one silicon drift detector (SDD) was mounted at $90^\circ$ (detector 1) and another SDD was mounted at $135^\circ$ (detector 2) with respect to the beam direction. The specification of the detectors (KETEK AXAS-A) are as follows: active area is 20 $mm^2$, thickness of Berrylium window is $8\mu m$ and FWHM is 124.2 ev for Mn $K_\alpha$ x-rays. The target ladder was mounted $90^\circ$ with respect to the beam direction. The target ladder could have five targets at a time and could be vertically displaced by a manual target manipulator. The distance between the target ladder and detectors was 14cm.  To measure the incident charge on target, one Faraday cup was mounted behind the target and attached to the current integrator. A collimator of 5 mm was placed inside the chamber in the line of beam entrance and 18 cm before the target ladder. Thin (thickness = $20\mu g/cm^2$) and spectroscopically pure target of Al was mounted on the target ladder. The thickness of the target was measured by energy loss method using $\alpha$-beam from $^{241}Am$ source.\\
\par The thickness of mylar window of chamber for detector 1 and detector 2 were $6\mu m$ and $10\mu m$, respectively.  The energy calibration was done before and after the experiment using $^{241}Am$ radioactive source. One position of the target of the target ladder was kept empty to take blank run. At another position of the target ladder a quartz crystal was placed to visualize the beam position. The online data  was acquired  using a  CAMAC based software called FREEDOM developed at IUAC \cite{freedomacquisition}.
\section{Results and discussion}
K x-ray spectra were recorded for Ne beam on Al-target at 1.8 and 2.1 Mev. Corresponding spectra have been shown in the Fig. 3. Fig. 3($C_1$) shows the raw spectrum containing a prominent Bremsstrahlung background. Subtracting the background from the raw spectrum we obtain the real nature of the spectrum. The back ground subtracted spectrum is shown in Fig. 3($D_1$). It shows only a good peak structure around 1.5 keV, which can be fitted into three Gaussian structures as shown with green lines. The same treatment is done on the raw spectrum obtained at 2.1 kev. Here, we seen the similar structure around 1.5 keV along with an additional structure around 3.0 keV. In order to check the ingenuity of the second structure we took a look on raw background spectrum, meaning the beam was passed through the target holder and without the target foil. The raw spectrum display exactly the same shape that we subtracted from the raw spectrum obtained with the target foils. It ensures the fact that the beam hallow hits the target frame. Now, similar back ground subtraction treatment done on the background spectrum displays a structure closed to 3 keV. Thus, such appearance questions on the ingenuity of the structure that pops up with the 2.1 MeV spectrum. \\
\begin{figure*}
\includegraphics[width=175mm,height=90mm,scale=01.0]{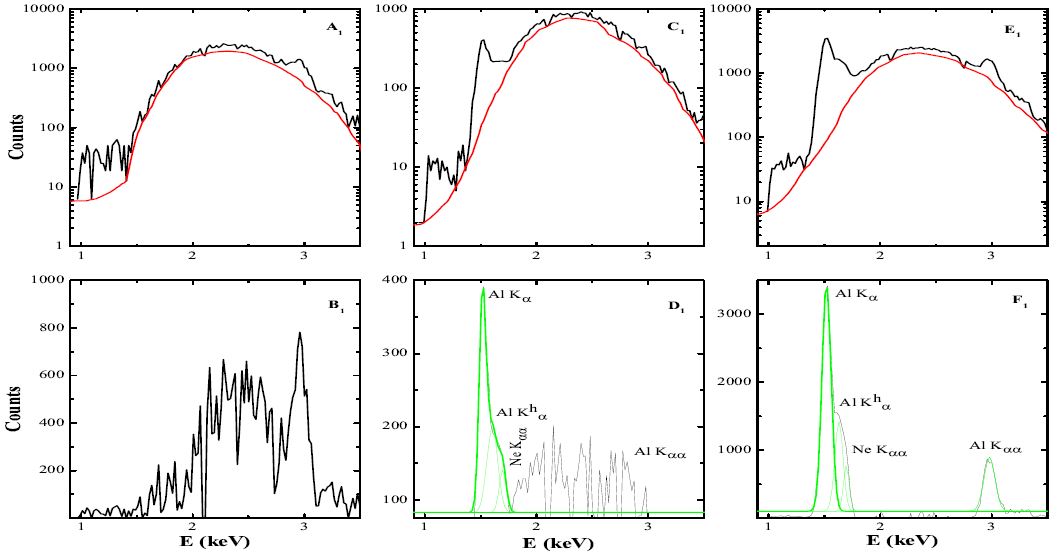}
\caption{\label{fig-1.8 MeV} ($A_1$) Line in black represent the spectrum of Ne on Al at 1.8 MeV energy and that in red the background spectrum, ($B_1$) the spectrum obtained after subtraction of the red spectrum from the black. The absolute spectrum so obtained is fitted with Gaussian peaks as shown with green curve, where E is the x-ray energy in keV and C the counts.($C_1$) Line in black represent the spectrum of Ne on Al at 2.1 MeV energy and that in red the background spectrum, ($D_1$) the spectrum obtained after subtraction of the red spectrum from the black. The absolute spectrum so obtained is fitted with Gaussian peaks as shown with green curve, where E is the x-ray energy in keV and C the counts. ($E_1$) Line in black represent the raw  background  spectrum at 2.1 MeV energy and that in red the Bremsstrahlung background spectrum, ($F_1$) the spectrum obtained after subtraction of the red spectrum from the black. The absolute spectrum so obtained is shown.}
\end{figure*}
\par This time instead of subtracting the background shown above, we subtract the actual background from the raw spectrum. This routine certainly deducted the contribution of the actual background from the spectrum and provided us the concrete shape of the spectrum. The correct spectrum so obtained does have a clear structure at 3.0 keV but much reduced intensity as expected. \\  

\begin{table*}[]
\tiny
\centering
\caption{{\label{tab:transition energy} Comparison of the transition energies (in keV) of  Ne and Al $K_\alpha$}, {$K^h_\alpha$} and {$K_{\alpha\alpha}$} peaks observed in this experiment vs earlier theories and experiments; where, El stands for element name, for this experiment $E_1$ = 1.8 MeV and $E_2$ = 2.1 MeV, A for theoretical values reported by others, B for experimental values reported by others, ND stands for not detected in this experiment, NA - not available in literature.} 
\begin{tabular}{|l|l|l|l|l|l|l|l|l|l|l|l|l|l|l|l|l|}
\hline
\multirow{2}{*}{\textbf{El}} & \multicolumn{4}{|l|}{\hspace{2cm}\textbf{$K_\alpha$}}     & \multicolumn{4}{|l|}{\hspace{2cm}\textbf{$K^h_\alpha$}}& \multicolumn{4}{|l|}{\hspace{2cm}\textbf{$K_{\alpha\alpha}$}}
\\ \cline{2-13} 
               & $E_1$ & $E_2$ & \textbf{\hspace{0.2cm}A} & {\textbf{\hspace{0.2cm} B}}& $E_1$ & $E_2$ & \textbf{\hspace{0.2cm}A} & {\textbf{\hspace{0.2cm} B}}& $E_1$ & $E_2$ & \textbf{\hspace{0.2cm}A} & {\textbf{\hspace{0.2cm} B}}\\ 
\hline

Ne & \makecell{ND} &\makecell{ND} & 0.848 \cite{winick2009x} & \makecell{NA} &\makecell{ND}&\makecell{ND} & \makecell{NA} &\makecell{NA}& \makecell{1.707 \\$\pm$ 0.207} & \makecell{1.725 \\$\pm$ 0.268}& \makecell{1.762 \cite{gavrila1976calculation}}& \makecell{1.8 \cite{andriamonje1991two}}   \\ \hline
Al & \makecell{1.521  \\$\pm 0.237$}&\makecell{1.518 \\$\pm$ 0.055} & 1.486 \cite{winick2009x} & \makecell{1.52\\$\pm$ 0.010 \cite{stoller1977two}}    &\makecell{1.637 \\$\pm$ 0.256}& \makecell{1.637 \\$\pm$ 0.117} & 1.622 \cite{saha2009effect} & \makecell{NA} &\makecell{ 3.002 \\$\pm$ 0.251}&\makecell{ 3.012 \\$\pm$ 0.194}& \makecell{3.072 \cite{kadrekar2010two},\\ 3.057 \cite{saha2009effect}} &\makecell{3.18\\ $\pm$ 0.020 \cite{stoller1977two}, \\\makecell{3.056\\$\pm$ 0.009 \cite{hoszowska2011first}}} \\ \hline
\end{tabular}
\end{table*}

\par We can see the the figures $3(D_1)$ and $3(F_1)$ display Al $K_\alpha$ peak (at 1.518 keV), $K^h_\alpha$ peak (at 1.637 keV), Ne $K_{\alpha\alpha}$ (at 1.725 keV) and Al $K_{\alpha\alpha}$ (at 3.012 KeV). These observed peaks have been compared well with the theoretical and experimental results as shown in Table \ref{tab:transition energy}.  The Table shows the {$K_\alpha$}, {$K^h_\alpha$} and {$K_{\alpha\alpha}$} energies of Ne and Al. Al $K_\alpha$, Al $K^h_\alpha$ and $K_{\alpha\alpha}$ transition energies are observed at 1.518 keV, 1.637 keV and 3.012 keV, respectively. While the Ne $K_{\alpha\alpha}$ transition energy is found 1.725 keV. It's an unusual observation as the Ne $K_{\alpha}$ as well as Ne $K^h_{\alpha}$ have not been observed but the Ne $K_{\alpha\alpha}$ is observed! This has happened due to the fact that much higher attenuation of $K_{\alpha}$ ($\approx$ 0.85 keV) than the $K_{\alpha\alpha}$ ($\approx$ 1.7 keV) as evidenced from Table \ref{tab:attenuation}. Further, the absolute efficiency of the SDD detector for  $K_{\alpha}$ is much less than that for the $K_{\alpha\alpha}$. \\
\par Making x-ray spectroscopy of light ions is very difficult, which can be understood from comparison of x-ray fluorescence yield as a function of atomic number \cite{chen1980relativistic,chen1991auger}. To visualise this fact we have compared the total K-level Auger transition rates ({$\Gamma_{KA}$} in unit of eV/$\hbar$), total K x-ray emission rate ({$\Gamma_{KF}$} in unit of eV/$\hbar$). Further, the percentage ratio of {$\Gamma_{KF}$} to {$\Gamma_{KA}$} for Ne and Al are given in Table \ref{tab:Auger-Flourescence}. We can see from the Table that value of Auger transition rate is much higher than the x-ray emission rate for K shell transitions. This ratio for Ne is only 1.89\%, this for Al is 5\%. 
\begin{table}
\caption{\label{tab:Auger-Flourescence} Auger transition rates ({${A_K}$}) and x-ray emission rates ({${R}$}) due to vacancy in K-shell (in unit of eV/$\hbar$) and percentage ratio of  {$R$} to {${A_K}$} for Ne and Al.}
\begin{tabular}{l l l l l l l l}
\hline
\makecell{Element}& \makecell{${A_K}$ \cite{mcguire1970k}}&\makecell{$R$ \cite{mcguire1970k}}&  \makecell{${R}$/${A_K}$}(\% )\\ 
\makecell{Ne} & \makecell{0.258}& \makecell{0.0049}& \makecell{1.89}\\
\makecell{Al} &\makecell{0.316}&  \makecell{0.0158} & \makecell{5}\\
\hline
\end{tabular}
\end{table}
\begin{table}
\caption{\label{tab:attenuation} Comparison of percentage attenuation of the intensity (\%) of Ne {$K_\alpha$} peak vs percentage attenuation of the intensity of Al {$K_\alpha$} peak in $6\mu m$ and $10\mu m$ mylar windows}
\begin{tabular}{l l l l l l l l}
\hline
\makecell{Thickness}& \makecell{Ne {$K_\alpha$}} &\makecell{Al {$K_\alpha$} } &  \makecell{Ne {$K_{\alpha\alpha}$} } &  \makecell{Al {$K_{\alpha\alpha}$}}\\ 
\makecell{ $6\mu m$} & \makecell{91.33}& \makecell{55.11} & \makecell{29.76} & \makecell{10.25} \\
\makecell{ $10\mu m$} &\makecell{98.30} & \makecell{73.68}& \makecell{44.50} & \makecell{16.50}\\
\hline
\end{tabular}
\end{table}
 
\begin{table*}
\caption{\label{tab:cross section} Production cross-sections ({${\sigma^x}_{K_\alpha}$}, {${\sigma^x}_{{K^h_\alpha}}$}, {${\sigma^x}_{K_{\alpha\alpha}}$} (in unit of mili-barn)) of the transitions {Al ${K_\alpha}$}, {Al {${K^h_{\alpha}}$}}, {Ne ${K_{\alpha\alpha}}$} and {Al ${K_{\alpha\alpha}}$}; the ratio of {Al {${K^h_{\alpha}}$}}, {Ne ${K_{\alpha\alpha}}$} and {Al ${K_{\alpha\alpha}}$} to the Al ${K_\alpha}$}
\begin{tabular}{l l l l l l l l l l l}
\hline
\makecell{E (MeV)}& \makecell{${{\sigma^x}_{K_\alpha}} (Al)$} & \makecell{{${\sigma^x}_{K{^h_{\alpha}}}$} (Al)} & \makecell{${{\sigma^x}_{K_{\alpha\alpha}}}$} (Ne)& \makecell{{${\sigma^x}_{K_{\alpha\alpha}}$} (Al)}& $\frac{{{\sigma^x}_{K{^h_{\alpha}}}} (Al)}{{{\sigma^x}_{K_\alpha}} (Al)}$ & $\frac{{\sigma^x}_{K_{\alpha\alpha}} (Ne)}{{{\sigma^x}_{K_\alpha}} (Al)}$ &$\frac{{\sigma^x}_{K_{\alpha\alpha}} (Al)}{{{\sigma^x}_{K_\alpha}} (Al)}$\\
\makecell{1.8} & \makecell{122}& \makecell{20}& \makecell{10} & \makecell{0.8}& \makecell{0.164} & \makecell{0.082} & \makecell{0.00656}\\
\makecell{2.1} &\makecell{694}& \makecell{82} & \makecell{25} & \makecell{3.6}&\makecell{0.118} &\makecell{0.036}& \makecell{0.00519} \\
\hline
\end{tabular}
\end{table*}

\begin{table*}
\caption{\label{tab:cross section} The value of absolute intensity (I: in the multiple of $10^6$) of the transitions {${K_\alpha}$}, {${{K^h_\alpha}}$}, {${K_{\alpha\alpha}}$; intensity ratios of {${{K^h_\alpha}}$}, {${K_{\alpha\alpha}}$} transitions to ${K_\alpha}$ transition}}
\begin{tabular}{l l l l l l l l l l l}
\hline
\makecell{E (MeV)}& \makecell{${{I_{K_\alpha}}} (Al)$} & \makecell{{${I_{K^h_{\alpha}}}$} (Al)} & \makecell{${{I_{K_{\alpha\alpha}}}} (Ne)$}& \makecell{{${I_{K_{\alpha\alpha}}}$} (Al)}& $\frac{I_{K^h_{{\alpha}}} (Al)}{{I_{K_{\alpha}}} (Al)}$ & $\frac{{I_{K_{\alpha\alpha}}} (Ne)}{{I_{K{_\alpha}}} (Al)}$& $\frac{{I_{K_{\alpha\alpha}}} (Al)}{{I_{K_{\alpha}}} (Al)}$\\
\makecell{1.8} & 
\makecell{203} &\makecell{33.2}& \makecell{16.8} & \makecell{1.45}& \makecell{$0.164$} &\makecell{$0.083$}& \makecell{$0.007$}\\
\makecell{2.1}&
\makecell{2420}& \makecell{285}& \makecell{85.7} & \makecell{12.6}& \makecell{$0.118$}& \makecell{$0.035$}& \makecell{$0.005$}\\
\hline
\end{tabular}
\end{table*}

\section{Summary}
To the best of our knowledge such a study is made for the first time in the country. We have succeeded to get the signature of TEOP transitions for the Ne and Al system at two low energy points (90 keV/amu and 95 keV/amu). In recorded spectrum of Ne beam on Al-target, we get a prominent Bremsstrahlung background. After subtraction of the Bremsstrahlung background from the raw spectrum, we get actual spectrum. We observed TEOP transition energies for Ne at 1.725 keV and Al at 3.012 keV. We also observed Al $K_\alpha$ and Al $K^h_\alpha$ peaks at 1.518 keV, 1.637 keV energies, respectively. Within experimental uncertainty limits, these values are well matched  with previously reported experimental and theoretical values. For both the projectile energies, observed values of transition energies are slightly different but within the measurement uncertainties.  
 
\section{Future Plans} 
\par We have presented significant results regarding the TEOP process, but it can only a start up. To study the TEOP process well in this laboratory, we have to upgrade our measurement system to great extent. In first instance, the beam focusing issue must be resolved. Beam need to be well focused and much smaller in in size that that of the foil, in our case it is 10 mm in diameter. We saw that little higher energy will be better to have higher statistics. We can not increase the deck potential more than 350 KV, hence we have to try to increase the charge state. Hope to get $Ne^{8+}$ for our use next time. We will put the detector closer to the target too. Further, measured production cross-section of the $K_{\alpha\alpha}$ transitions are extremely useful to test the theoretical iedeas relevant to the TEOP processes. Note that x-ray production cross sections for the K x-ray lines can be determined from the following formula
\begin{equation}
\sigma_i^x=(Y_i^x A Sin\theta)/(N_A\epsilon n_p t \beta) \label{eqn:3}
\end{equation}
\noindent where $Y_i^x$ is the intensity of the ith x-ray peak ($i=K_\alpha,K_\beta$), A is the atomic weight of the target, $\theta$ is the angle between the incident ion beam and the target foil normal, $N_A$ is the Avogadro number, $n_p$ is the number of incident projectiles, $\epsilon$ is the effective efficiency of the x-ray detector, t is the target thickness in $\mu g/cm^2$ and $\beta$ is a correction factor for energy loss of the incident projectile and absorption of emitted x-rays in the target element. Nevertheless, we have not yet measured the absolute efficiency of the SDD detector. Hence, we may determine only the relative cross section of  $K_{\alpha\alpha}$ with respect to  $K_{\alpha}$ and  $K^h_{\alpha}$ from the relative intensity ratios if we know the relative detector efficiency from any source, for example \citet{schlosser2010expanding}. Though this procedure is well applicable to Al, but not so for Ne at all as we have not detected the Ne  $K_{\alpha\alpha}$ and  $K^h_{\alpha}$ transitions. We can think of two remedies for this issue. Number one is to place the SDD detector inside the vacuum chamber and number two is to measure the quantum efficiency of the detector. It appears the first option is too difficult to to adapt for the SDD detector of our disposal. Hence, we plan to measure the over all detector efficiency (including solid angle, mylar window and air gap between the detector and mylar window etc.) by using the proton beam of 350 keV and the targets possibly from Mg to Ge.\\
\par It is well known that for the light elements, rate of Auger transition is much higher than fluorescence rate. So we will add an experimental set up for electron spectroscopy in this experiment in future which will give us higher statistics and thus can understand better the two electron and one photon phenomenon. Note that the electron spectroscopy has not yet been applied for the study of two electron and one photon processes. 

\begin{acknowledgments}
One of the authors, Shashank Singh is thankful to the staff of LEIBF facility of IUAC (New Delhi, India) for providing excellent beam and necessary facilities for the experiment. 
\end{acknowledgments}

\nocite{*}

\bibliography{apssamp}

\end{document}